\documentclass{ifacconf}
\usepackage{graphicx}      % include this line if your document contains figures
\usepackage{natbib}        % required for bibliography
\usepackage{times} % assumes new font selection scheme installed
\usepackage{amsmath} % assumes amsmath package installed
\usepackage{amssymb}  % assumes amsmath package installed
\def\sn{|\!|\!|}

\newcommand{\sinc}{\mathrm{sinc}}
\newcommand{\tanc}{\mathrm{tanc}}
\newcommand{\tanhc}{\mathrm{tanhc}}

\DeclareMathAlphabet{\bit}{OML}{cmm}{b}{it}
\def\<{\leqslant}           % nice less than or equal to sign
\def\>{\geqslant}           % nice larger than or equal to sign
\def\d{\partial}

\def\Re{\mathrm{Re} }   % real part
\def\Im{\mathrm{Im} }   % imaginary part
\def\rprod{\mathop{\overrightarrow{\prod}}}

\def\mZ{{\mathbb Z}}    % set of integers
\def\mR{{\mathbb R}}    % real line
\def\mC{\mathbb{C}}    % complex plane

\def\Tr{\mathrm{Tr}}       % matrix trace
\def\rT{{\rm T}}        % matrix transpose
\def\diam{\diamond}       % matrix trace
\def\bP{\mathbf{P}}    % probability
\def\bE{\mathbf{E}}    % expectation
\def\bM{\mathbf{M}}    % classical expectation

\def\[[[{[\![\![}
\def\]]]{]\!]\!]}

\def\dbra{\langle\!\langle}
\def\dket{\rangle\!\rangle}

\def\re{{\rm e}}        % number e
\def\rd{{\rm d}}        % differential

\def\bJ{\mathbf{J}}

\def\br{\mathbf{r}}
\def\x{\times}
\def\ox{\otimes}

\def\fB{\mathfrak{B}}

\def\fF{\mathfrak{F}}

\def\fH{\mathfrak{H}}

\def\cW{\mathcal{W}}

\def\cI{\mathcal{I}}

\def\cov{\mathbf{cov}}

\def\mZ{{\mathbb Z}}

\def\eps{\epsilon}
\def\ups{\upsilon}
\def\Ups{\Upsilon}

\def\diag{\mathop\mathrm{diag}}    % diagonal matrix
\begin{document}
\begin{frontmatter}

% in
%Tail Distribution Bounds for Nonlinear Quantum Stochastic Systems using Quadratic-Exponential Moments

\title{Probabilistic Bounds with Quadratic-Exponential Moments for
Quantum Stochastic Systems\thanksref{footnoteinfo}}

\thanks[footnoteinfo]{This work is supported by the Australian Research Council  grants  DP210101938, DP200102945.}

\author[IGV]{Igor G. Vladimirov}%${}^*$, \quad Ian R. Petersen}

\address[IGV]{School of Engineering, Australian National University, ACT 2601, Canberra, Australia    (e-mail: igor.g.vladimirov@gmail.com).}

\begin{abstract}                % Abstract of not more than 250 words.
This paper is concerned with quadratic-exponential moments (QEMs) for  dynamic variables of  quantum stochastic systems with position-momentum type canonical commutation relations. The QEMs play an important  role for statistical ``localisation'' of the quantum dynamics in the form of upper bounds on the tail probability distribution for a positive definite quadratic function of the system variables.
We employ a randomised representation of the QEMs in terms of the moment-generating function (MGF) of the system variables, which is averaged over its parameters using an auxiliary classical  Gaussian random vector. This representation is combined with a family of weighted $L^2$-norms of the MGF, leading to upper bounds for the QEMs of the system variables. These bounds are demonstrated for open quantum harmonic oscillators with vacuum input fields and non-Gaussian initial states.
\end{abstract}


\end{frontmatter}

\section{Introduction}\label{sec:intro}

Quantum mechanical models, which describe light-matter interaction on an atomic scale,  often originate as a quantisation of classical systems  and have practical applications in quantum optics, quantum metrology, quantum communication and quantum information processing (\cite{NC_2000,WM_2008,WM_2010}). These applications  exploit quantum mechanical resources beyond classical (deterministic or stochastic) dynamics to engineer quantum systems with enhanced performance (involving stability, optimality and robustness requirements).
Similarly to classical counterparts, in the case of quantum systems,
it is important to have tools for obtaining upper bounds on cost functionals (such as mixed moments of the system variables and probabilities of relevant events)  which would allow their behaviour to be statistically ``localised'',  with a high degree of confidence, to a bounded domain. The need in such localisation can arise,  for example, in determining the range of applicability of a linearised  model which results from neglecting nonlinearities in the actual quantum dynamics, provided the terms being omitted are small enough in a suitable sense.

However, a distinctive  feature of quantum probability (\cite{H_2001}) is that, unlike real-valued  physical quantities, the   dynamic variables of quantum mechanical systems are linear operators on a Hilbert space, whose statistical properties (such as mean values) are specified in terms of operator-valued quantum states (or density operators on the same space). Because of these more complicated operator-valued structures (also playing a role in the interpretation of quantum measurement  as an interaction of a quantum system with a classical device modifying  the quantum state),  noncommuting quantum variables cannot have classical joint probability distributions. 
Nevertheless, any single self-adjoint quantum variable has a well-defined probability distribution on the real line, obtained by averaging the projection-valued measure of this operator. Moreover, pairwise commuting self-adjoint quantum variables are isomorphic to a set of classical real-valued random variables on a common probability space with Kolmogorov's axiomatics (\cite{S_1996})  and thus have not only marginal distributions, but also  a joint probability distribution in an appropriately dimensioned  Euclidean space.

The present paper is concerned with quantum stochastic systems whose dynamic variables have a position-momentum type commutation structure, typical for open quantum harmonic and anharmonic    oscillators interacting with external bosonic fields. In the framework of the Hudson-Parthasarathy calculus (\cite{HP_1984,P_1992}),  such systems are  modelled by quantum stochastic differential equations (QSDEs) driven by quantum Wiener processes on a symmetric Fock space. We discuss an approach to statistical ``localisation'' in terms of the tail probability distribution for a self-adjoint operator-valued positive definite quadratic function of the system variables. Efficient upper bounds for the tail probabilities are provided by using quadratic-exponential moments (QEMs)  of the system variables. Being similar to the cost functionals in classical risk-sensitive control (\cite{J_1973,W_1981}), the QEMs and related structures play an important  role in quantum control (\cite{B_1996,VPJ_2021}), operator algebras (\cite{AB_2018}), quantum L\'{e}vy area problems (\cite{H_2018}), and  equilibrium statistical mechanical computations for Gaussian quantum states (\cite{PS_2015}). While the QEMs lend themselves to closed-form calculation for Gaussian states, the situation is different in the non-Gaussian case, typical for nonlinear quantum systems,  where direct analysis of the dynamics of such moments is also complicated (\cite{VP_2012}). As an alternative route, we employ a randomised representation of the QEMs, which, similarly to (\cite{VPJ_2019}), is applicable to general (not only Gaussian) quantum states and involves a classical   averaging  of the moment-generating function (MGF) of the quantum system variables using an auxiliary Gaussian random vector. This allows upper bounds on the QEMs (and the tail probabilities) to be obtained from those for the MGF, with the latter being more tractable. More precisely, the QEM bounds, established in the present paper, involve a family of weighted $L^2$-norms of the MGF from (\cite{V_2015}). These bounds are demonstrated for open quantum harmonic oscillators (OQHOs), governed by linear QSDEs with vacuum input fields and non-Gaussian initial states.

The paper is organised as follows.
Section~\ref{sec:var} specifies the class of quantum variables of position-momentum type being
considered.
Section~\ref{sec:quadro} describes the QEMs and reviews their role for upper bounds on the tail probabilities.
Section~\ref{sec:rand} provides a randomised representation for the QEMs of the system variables in terms of their MGF.
Section~\ref{sec:Gauss} calculates the QEMs for Gaussian states and establishes upper bounds  for the QEMs in terms of the weighted $L^2$-norms of the MGF in the non-Gaussian case.
Section~\ref{sec:OQHO} demonstrates the QEM bounds for OQHOs with vacuum input fields and non-Gaussian initial states.
Section~\ref{sec:conc} makes concluding remarks.
Appendix~\ref{sec:MMM} provides an alternative proof regarding the randomised QEM representation in the classical case.

\section{Quantum variables with position-momentum type commutations}\label{sec:var}

As counterparts of classical Hamiltonian systems (\cite{A_1989}) with conjugate position-momentum pairs, continuous variables quantum systems are  usually endowed with  an even number
\begin{equation}
\label{nnu}
      n := 2\nu
\end{equation}
of dynamic variables $X_1, \ldots, X_n$ which are self-adjoint operators on a complex separable Hilbert space $\fH$ with the Heisenberg canonical commutation relations (CCRs)
\begin{equation}
\label{XCCR}
    [X,X^\rT] := ([X_j,X_k])_{1\< j,k\< n} = 2i \Theta
\end{equation}
(on a dense domain of $\fH$) specified by a real antisymmetric matrix  $\Theta = -\Theta^\rT \in \mR^{n \x n}$. Here, the system variables are assembled into a column-vector $X:= (X_k)_{1\< k\< n}$,    the transpose $(\cdot)^\rT$ acts on vectors and matrices of operators as if their entries were scalars, $[\alpha,\beta]:= \alpha\beta - \beta\alpha$ is the  commutator of linear operators, and $i:= \sqrt{-1}$ is the imaginary unit.   The relation (\ref{XCCR}) is an infinitesimal form of the Weyl CCRs (\cite{F_1989})
\begin{equation}
\label{Weyl}
  \cW(u+v)
  =
  \re^{iu^\rT\Theta v}
  \cW(u) \cW(v),
      \qquad
      u,v\in \mR^\nu
\end{equation}
(whereby $[\cW(u), \cW(v)] = -2i\sin(u^\rT\Theta v)\cW(u+v)$)
for the unitary Weyl operators
\begin{equation}
\label{cW}
  \cW(u):= \re^{iu^\rT X},
  \qquad
  u \in \mR^n,
\end{equation}
associated with the system variables. The CCRs (\ref{XCCR}), (\ref{Weyl}) can be obtained by using linear combinations of another set of quantum variables with a simpler commutation structure, as described below for completeness.
To this end, the CCR matrix $\Theta$ in (\ref{XCCR}) is assumed to be nonsingular:    $\det \Theta \ne 0$, and hence, $\Theta$ has purely  imaginary eigenvalues $\pm i \theta_k\ne 0$ with the corresponding eigenvectors $w_k, \overline{w_k} \in \mC^n$:
\begin{equation}
\label{eig}
    \Theta
    \begin{bmatrix}
      w_k & \overline{w_k}
    \end{bmatrix}
    =
    i \theta_k
    \begin{bmatrix}
      w_k & \overline{w_k}
    \end{bmatrix}
    \begin{bmatrix}
      1 & 0 \\
      0 & -1
    \end{bmatrix},
  \qquad
  k = 1, \ldots, \nu,
\end{equation}
where $\overline{(\cdot)}$ is the complex conjugate, and $\theta_k >0$ are the eigenfrequencies of $\Theta$ (so that $\pm\theta_k$ are the eigenvalues of the Hermitian matrix $-i\Theta$). The eigenvectors of $\Theta$ can be chosen so as to form an orthonormal basis in $\mC^n$: 
\begin{equation}
\label{ort}
    \begin{bmatrix}
      w_j & \overline{w_j}
    \end{bmatrix}^*
    \begin{bmatrix}
      w_k & \overline{w_k}
    \end{bmatrix}
    =
    \begin{bmatrix}
      w_j^* w_k & w_j^*\overline{w_k}\\
      w_j^\rT w_k & w_j^\rT \overline{w_k}
    \end{bmatrix}
    =
    \delta_{jk}
    I_2
\end{equation}
for all $  j, k = 1, \ldots, \nu$,
where $(\cdot)^*:= \overline{(\cdot)}{}^\rT$ is the complex conjugate transpose,
$\delta_{jk}$ is the Kronecker delta, and $I_r$ is the identity matrix of order $r$. In terms of the real and imaginary parts
\begin{equation}
\label{uvw}
  u_k := \Re w_k,
  \qquad
  v_k := \Im w_k
\end{equation}
of the eigenvectors, the relations (\ref{eig}), (\ref{ort}) are represented  as
\begin{equation}
\label{eig1_ort1}
    \Theta
    h_k
    =
    \theta_k
    h_k
    \bJ,
    \qquad
    h_j^\rT h_k
    =
    \begin{bmatrix}
      u_j^\rT u_k & u_j^\rT v_k\\
      v_j^\rT u_k & v_j^\rT v_k
    \end{bmatrix}
    =
    \frac{1}{2}
    \delta_{jk}I_2,
\end{equation}
where
\begin{equation}
\label{huv}
  h_k
  :=
      \begin{bmatrix}
      u_k & v_k
    \end{bmatrix}
    \in \mR^{n\x 2},
\end{equation}
and the matrix
\begin{equation}
\label{bJ}
\bJ: = {\begin{bmatrix}
        0 & 1\\
        -1 & 0
    \end{bmatrix}}
\end{equation}
spans the subspace of antisymmetric matrices of order 2. In turn, (\ref{eig1_ort1}) is equivalent to
\begin{equation}
\label{eig2_ort2}
    \Theta
    H
    =
    H
    (
        \Gamma
    \ox \bJ
    ),
    \qquad
    H^\rT H = (h_j^\rT h_k)_{1\< j,k\< \nu} = \frac{1}{2} I_n
\end{equation}
(with $\ox$ the Kronecker product of matrices), where
\begin{equation}
\label{diag}
    \Gamma:=
      \diag_{1\< k \< \nu}
    (\theta_k),
\end{equation}
and
\begin{equation}
\label{Hhh}
    H
    :=
    \begin{bmatrix}
      h_1 & \ldots & h_{\nu}
    \end{bmatrix}
    =
    \begin{bmatrix}
      u_1 & v_1 & \ldots & u_{\nu} & v_{\nu}
    \end{bmatrix}
    \in \mR^{n\x n},
\end{equation}
so that 
the matrix $\sqrt{2}H$ is orthogonal, and hence, (\ref{eig2_ort2}) yields
\begin{equation}
\label{TH}
    \Theta
    =
    2H
    (
        \Gamma
    \ox \bJ
    )
    H^\rT
    =
    2\sum_{k=1}^\nu
    \theta_k h_k \bJ h_k^\rT.
\end{equation}
If $f$ is a function of a complex variable, holomorphic in a neighbourhood of the spectrum $\{\pm i \theta_k: k = 1, \ldots, \nu\}$ of the matrix $\Theta$,
then the matrix $f(\Theta) \in \mC^{n\x n}$ has the eigenvalues $f(\pm i\theta_k)$, inherits  from $\Theta$ the eigenvectors $w_k$, $\overline{w_k}$ in (\ref{eig}) and, in view of (\ref{TH}), can be represented as
\begin{align}
\nonumber
    f(\Theta)
    & =
    2Hf(\Gamma \ox \bJ) H^\rT\\
\label{phiT}
    & =
        2H(f_+(i\Gamma) \ox I_2 - if_-(i\Gamma)\ox \bJ) H^\rT,
\end{align}
where $f_{\pm}(z):=\frac{1}{2}(f(z)\pm f(-z))$ are the symmetric and antisymmetric  parts of $f$, so that $f(\pm z) = f_+(z)\pm f_-(z)$. In particular, if $f$ is symmetric ($f_-\equiv 0$),  then (\ref{phiT}) leads to
\begin{equation}
\label{symphiT_phiTtrace}
    f(\Theta)
    =
        2H(f(i\Gamma) \ox I_2) H^\rT,
        \quad
            \Tr f(\Theta)
    =
    2\sum_{k=1}^{\nu} f(i\theta_k). \!
\end{equation}
Now, with the system variables $X_1, \ldots, X_n$, we
associate the following quantum variables on the Hilbert space $\fH$:
\begin{align}
\label{ak}
  a_k
  & :=
  \frac{1}{\sqrt{\theta_k}}
  w_k^\rT
  X
  =
  q_k + ip_k,\\
\label{qk}
    q_k
    & :=
    \Re
    a_k
    =
  \frac{1}{\sqrt{\theta_k}}
  u_k^\rT   X,\\
\label{pk}
    p_k
    & :=
    \Im
    a_k
    =
  \frac{1}{\sqrt{\theta_k}}
  v_k^\rT X,
    \qquad
   k = 1, \ldots, \nu.
\end{align}
where
the orthonormal eigenvectors (\ref{eig}) of the CCR matrix $\Theta$ from (\ref{XCCR}) are used together with (\ref{uvw}) and  the eigenfrequencies $\theta_k$. Here, the real and imaginary parts are extended from the complex plane  $\mC$
to quantum variables as
$\Re z := \frac{1}{2} (z + z^\dagger)$,
$\Im z := \frac{1}{2i} (z - z^\dagger)$, where $(\cdot)^\dagger$ is the operator adjoint. Accordingly,
\begin{equation}
\label{zetk+}
  a_k^\dagger
  =
  \frac{1}{\sqrt{\theta_k}}
  w_k^* X
  =
  q_k - ip_k,
\end{equation}
in view of (\ref{ak}) and self-adjointness of the quantum variables $X_1, \ldots, X_n$ and $q_k$, $p_k$ in (\ref{qk}), (\ref{pk}). By using (\ref{Hhh}), the relations (\ref{qk}), (\ref{pk}) can be  assembled into
\begin{equation}
\label{r_rk}
    r:=
    (r_k)_{1\< k \< \nu}
    =
    (
    \Gamma^{-1/2}
    \ox
    I_2
    )
    H^\rT X,
    \quad
  r_k
  :=
  {\begin{bmatrix}
    q_k\\
    p_k
  \end{bmatrix}},
\end{equation}
where $\Gamma^{-1/2} = \diag_{1\< k\< \nu}
    \big(1/\sqrt{\theta_k}\big)$ is associated with the eigenfrequencies of the CCR matrix $\Theta$ in view of (\ref{diag}).

\begin{lem}
\label{lem:posmom}
The quantum variables $a_k$, associated by (\ref{ak})--(\ref{pk}) with the quantum variables $X_1, \ldots, X_n$ and the eigenbasis of the nonsingular CCR matrix $\Theta$ in (\ref{eig}), (\ref{uvw}), satisfy
\begin{equation}
\label{acomm}
  [a_j, a_k^\dagger] = 2\delta_{jk},
  \qquad
  [a_j, a_k] = 0,
  \qquad
  [a_j^\dagger, a_k^\dagger] = 0.
\end{equation}
\end{lem}
\begin{pf}
By using (\ref{XCCR}), (\ref{eig}), (\ref{ort}), it follows from (\ref{ak}), (\ref{zetk+}) that
$    [a_j,a_k^\dagger]
    =
  \frac{1}{\sqrt{\theta_j\theta_k}}
  w_j^\rT
  [X,X^\rT]
  \overline{w_k}
  =
  \frac{2i}{\sqrt{\theta_j\theta_k}}
  w_j^\rT
  \Theta
  \overline{w_k}
  =
  2 \sqrt{\frac{\theta_k}{\theta_j}}
  w_j^\rT \overline{w_k}
 =
  2\delta_{jk}
$
for all $j,k=1, \ldots, \nu$, which establishes the first equality in (\ref{acomm}).
By a similar reasoning,
$
    [a_j,a_k]
    =
  \frac{1}{\sqrt{\theta_j\theta_k}}
  w_j^\rT
  [X,X^\rT]
  w_k
  =
  \frac{2i}{\sqrt{\theta_j\theta_k}}
  w_j^\rT
  \Theta
  w_k
  =
  -2 \sqrt{\frac{\theta_k}{\theta_j}}
  w_j^\rT w_k
 =
 0
$,
and hence,
$    [a_j^\dagger,a_k^\dagger]
    =
     -
    [a_j, a_k]^\dagger
    =0
$
for all $j,k = 1, \ldots, \nu$,
thus completing the proof of (\ref{acomm}).\hfill$\blacksquare$
\end{pf}

The CCRs (\ref{acomm}) show that $a_k$ in (\ref{ak}) are organised as  pairwise commuting annihilation operators (\cite{S_1994}), so that $a_k^\dagger$ in (\ref{zetk+}) are the corresponding  creation operators, and the self-adjoint quantum variables $q_k$, $p_k$ in (\ref{qk}), (\ref{pk}) form conjugate position-momentum pairs, with
\begin{equation}
\label{qpcomm}
  [q_j,q_k] = 0,
  \qquad
  [p_j,p_k] = 0,
  \qquad
  [q_j,p_k] = i \delta_{jk}.
\end{equation}
Accordingly, the vectors $r_k$ in (\ref{r_rk}) commute with each other (for different $k$) and have a common CCR matrix $\frac{1}{2}\bJ$:
\begin{equation}
\label{rrcomm}
    [r_j,r_k^\rT]
    =
    \begin{bmatrix}
      [q_j,q_k] & [q_j,p_k]\\
      [p_j,q_k] & [p_j,p_k]
    \end{bmatrix}
    =
    i \delta_{jk}
    \bJ,
\end{equation}
where (\ref{qpcomm}) is used together with (\ref{bJ}). The orthonormality (\ref{ort}) of the  eigenbasis in (\ref{eig}) allows the quantum variables $X_1, \ldots, X_n$ to be recovered from (\ref{ak}), (\ref{zetk+}) as
\begin{align}
\nonumber
    X
    &=
    \sum_{k=1}^{\nu}
    \sqrt{\theta_k}
    (\overline{w_k}
    a_k
    +
    w_k
    a_k^\dagger
    ) =
    2
    \sum_{k=1}^{\nu}
    \sqrt{\theta_k}
    \Re
    (    w_k
    a_k^\dagger
    )\\
\nonumber
    & =
    2
    \sum_{k=1}^{\nu}
    \sqrt{\theta_k}
    (u_kq_k + v_kp_k)
    =
    2
    \sum_{k=1}^{\nu}
    \sqrt{\theta_k}
    h_k
    r_k\\
\label{Xqp}
    & =
    2
    H
    (
    \sqrt{\Gamma}
    \ox I_2
    )
    r,
\end{align}
where  use is made of  (\ref{huv}), (\ref{r_rk}), and $\sqrt{\Gamma}= \diag_{1\< k \< \nu}
    (\sqrt{\theta_k})$ in view of (\ref{diag}). The last equality in (\ref{Xqp}) can also be obtained directly from (\ref{r_rk}) since
        $((
    \Gamma^{-1/2}
    \ox
    I_2
    )
    H^\rT)^{-1}
    = \sqrt{2}(\sqrt{2}H^\rT)^{-1} (
    \sqrt{\Gamma}
    \ox
    I_2
    )=2H(
    \sqrt{\Gamma}
    \ox
    I_2
    )$ due to orthogonality $\sqrt{2} H^\rT =\frac{1}{\sqrt{2}} H^{-1}$ of the matrix $\sqrt{2}H$. This will play a part in considering quadratic forms of $X$ in Sections~\ref{sec:quadro}, \ref{sec:rand}.

\section{Tail Distribution Bounds and Quadratic-Exponential Moments}
\label{sec:quadro}

Being noncommutative, the quantum system variables $X_1, \ldots,\\ X_n$ 
do not have a classical joint probability distribution in $\mR^n$ (\cite{H_2001}), and their behaviour cannot be ``localised''  in terms of nonexisting probabilities $\bP(X \in B)$ for bounded Borel subsets of $\mR^n$.  Nevertheless, if $\xi_k := f_k(r_k)$ are self-adjoint quantum variables,  which are functions of the corresponding  position-momentum pairs (\ref{r_rk}) from (\ref{qk}), (\ref{pk}), then $[\xi_j,\xi_k] = 0$ for all $j,k =1, \ldots, \nu$ in view of (\ref{rrcomm}). Hence, $\xi:= (\xi_k)_{1\< k \< \nu}$ is isomorphic to a classical  $\mR^\nu$-valued random vector with a well-defined distribution $D(B):= \bP(\xi \in B)$ on the $\sigma$-algebra $\fB^\nu$ of Borel subsets of $\mR^\nu$, with the pushforward measure $D\circ g^{-1}$ describing the distribution of  the random variable $\zeta:= g(\xi)$ for any Borel function $g: \mR^\nu \to \mR$. Moreover, an arbitrary self-adjoint quantum variable $\zeta$ on the underlying Hilbert space $\fH$ (which can be a more general  function of the system variables $X_1, \ldots, X_n$, not reducible to that of $\xi_1, \ldots, \xi_\nu$) has a classical distribution on $\mR$. The probability
\begin{equation}
\label{bP}
    \bP(\zeta \in A) := \bE P(A),
    \qquad
    A \in \fB,
\end{equation}
is defined in terms of the averaged projection-valued measure $P$ in the spectral decomposition  $\zeta = \int_\mR z P(\rd z)$   (satisfying $P(A\bigcap B) = P(A)P(B)$ for all $A, B \in \fB$, with $P(\mR) = \cI_\fH$ the identity operator on $\fH$). Here,
\begin{equation}
\label{bE}
    \bE \varphi
    :=
    \Tr(\rho \varphi)
\end{equation}
is the quantum expectation of an operator $\varphi$ on the space $\fH$  over an underlying density operator $\rho = \rho^\dagger \succcurlyeq 0$ on $\fH$ of unit  trace $\Tr \rho = 1$. Accordingly, the probability distribution of $\zeta$, defined by (\ref{bP}), satisfies the classical probabilistic inequalities (\cite{S_1996}), including
\begin{equation}
\label{tail0}
    \ln \bP(\zeta \> z)
    \<
    -
    \sup_{\mu >0}
    (
        z \mu
        -
        \ln \bE \re^{\mu \zeta}
    ),
    \qquad
    z\in \mR,
\end{equation}
where the supremum is the Legendre transformation (\cite{R_1970}) of the cumulant-generating function (CGF) $\mu \mapsto \ln \bE \re^{\mu \zeta}$ for the quantum variable $\zeta$. In the context of ``localising'' the statistical behaviour of $X_1, \ldots, X_n$
(in the absence of a joint distribution in the classical sense),
the tail distribution bounds (\ref{tail0}) are applicable to the positive semi-definite self-adjoint quantum variable
\begin{equation}
\label{QXX}
    Q := \sum_{k=1}^n X_k^2
    =
  X^\rT X
\end{equation}
(a quantum counterpart of the square of the  Euclidean norm $|\cdot|$ in $\mR^n$)
as
\begin{equation}
\label{tail}
    \bP(Q \> 2\eps)
    \<
    \re^{-
    \sup_{\mu >0}
    (
        \eps \mu
        -
        \Ups(\mu)
    )},
    \qquad
    \eps \> 0.
\end{equation}
Here,
\begin{equation}
\label{Ups}
    \Ups(\mu)
    :=
    \ln \Xi(\mu)
\end{equation}
is the CGF for the quantum variable $\frac{1}{2}Q$, related to the quadratic-exponential moments (QEMs)
\begin{equation}
\label{Xi}
  \Xi(\mu):= \bE \re^{\frac{\mu}{2}Q},
  \qquad
  \mu \> 0,
\end{equation}
of the quantum variables $X_1, \ldots, X_n$, with $\mu$ playing the role of a risk sensitivity parameter. Note that $\Xi(\mu)\> 1$ for any $\mu \> 0$ since $Q\succcurlyeq 0$,   and hence, $\Ups(\mu)\> 0$.
An alternative form of (\ref{tail}) is
$    \ln \bP(Q \> 2\Ups'(\mu))
    \<
    \Ups(\mu)-\mu\Ups'(\mu)$ for all $
    \mu > 0$,
where the right-hand side is the negative of the Bregman divergence (\cite{B_1967})  associated with the CGF $\Ups$ at the points $0$, $\mu$ in view of $\Ups(0)=0$.

Instead of $Q$ in (\ref{QXX}),  a more general quadratic form $X^\rT \Pi X$, with a positive  definite matrix $\Pi=\Pi^\rT \in \mR^{n\x n}$, can also be used. However, it is reduced to (\ref{QXX}) by replacing $X$ with its weighted version $\sqrt{\Pi} X$,   which, in comparison with (\ref{XCCR}), has an appropriately modified nonsingular CCR matrix $\sqrt{\Pi} \Theta  \sqrt{\Pi}$.

In the limiting case of commuting system variables $X_1, \ldots, X_n$ (with zero CCR matrix $\Theta = 0$), when  $X$ is a classical $\mR^n$-valued random vector,  there is an integral relation
\begin{align}
\nonumber
    \bM \re^{\frac{\mu}{2} |X|^2}
    & =
    \bM \bM(\re^{\sqrt{\mu}Z^\rT X}\mid X)
    =
        \bM \re^{\sqrt{\mu}Z^\rT X}\\
\nonumber
    & =
    \bM \bM(\re^{\sqrt{\mu}Z^\rT X}\mid Z)
    =
    \bM\Lambda(\sqrt{\mu}Z)\\
\label{MMM}
    & =
    (2\pi)^{-\nu}
    \int_{\mR^n}
    \re^{-\frac{1}{2}|z|^2}
    \Lambda(\sqrt{\mu}z)
    \rd z
\end{align}
between the QEMs and the moment-generating function (MGF)
\begin{equation}
\label{MGFclass}
  \Lambda(u):= \bM \re^{u^\rT X},
  \qquad
  u \in \mR^n,
\end{equation}
where $\bM(\cdot)$ is the classical expectation, with its conditional version $\bM(\cdot \mid \cdot)$. Here, $Z$ is an auxiliary standard normal random vector in $\mR^n$ (that is, Gaussian with zero mean $\bM Z = 0$ and the identity covariance matrix $\bM(ZZ^\rT) = I_n$), independent of $X$, so that the conditional distribution of $Z^\rT X$, given $X$, is Gaussian with zero mean $\bM(Z^\rT X \mid X) = X^\rT\bM Z = 0$  and variance $\bM((Z^\rT X)^2 \mid X) = X^\rT \bM(ZZ^\rT) X = |X|^2$. Hence, 
    $\bM (\re^{\sqrt{\mu}Z^\rT X} \mid X) = \re^{\frac{\mu}{2}|X|^2}$, 
which gives rise to the first equality in (\ref{MMM}).     The tower property of iterated conditional expectations (\cite{S_1996})  is then repeatedly employed in order to arrive at the averaging of the classical MGF $\Lambda$ from (\ref{MGFclass}) over its argument on the right-hand side of (\ref{MMM}).
The relation (\ref{MMM})  allows upper bounds on the MGF $\Lambda$ to be ``translated'' to those for the QEM of the classical random vector $X$ and the tail distributions (\ref{tail}).
However, the reasoning, based on classical conditional expectations as in (\ref{MMM}) (or a more elementary completion-of-the-square argument  in Appendix~\ref{sec:MMM}) uses the commutativity between $X_1, \ldots, X_n$ and   is not applicable in the noncommutative quantum case. Nevertheless,  the general randomization  idea can be retained, similarly to (\cite{VPJ_2019}),  with an appropriate modification, as discussed in the next section.

\section{A Randomised representation of QEMs}
\label{sec:rand}

We will first represent the quadratic-exponen\-ti\-al function $\re^{\frac{\mu}{2}Q}$ of $X$ in the QEM (\ref{Xi}) in a form which does not depend on a particular quantum state $\rho$ from (\ref{bE}). For this purpose, the following lemma  on ``elementary'' quadratic-exponential  functions will be used.

\begin{lem}
\label{lem:fact}\cite[Theorem A.1]{VPJ_2021}
For any $\omega\>0$, a quantum mechanical posi\-tion-momentum pair $(q,p)$ (with $[q,p] = i$) satisfies the identity
\begin{align}
\nonumber
    &\re^{\omega (q^2+p^2)} \cosh \omega
    =
  \bM \re^{\sigma(\alpha q + \beta p )}\\
\label{qefrand}
  & =
  \frac{1}{2\pi}
  \int_{\mR^2}
    \re^{\sigma(a q + bp )
    -\frac{1}{2}(a^2 +b^2)}
  \rd a \rd b,
    \ \
    \sigma := \sqrt{2\tanh \omega},
\end{align}
where the classical expectation  $\bM(\cdot)$ is over independent standard normal  random variables $\alpha$, $\beta$.
\hfill$\blacksquare$
\end{lem}

This lemma will be combined below with the representation (\ref{Xqp}) of the quantum variables $X_1, \ldots, X_n$ in terms of the position-momentum pairs $q_1, p_1, \ldots, q_\nu, p_\nu$ in (\ref{qk}), (\ref{pk}). To this end, (\ref{qefrand}) extends to
\begin{equation}
\label{qefrand1}
    \re^{\sum_{k=1}^\nu\omega_k s_k}
    =
    \frac{\bM \re^{\sum_{k=1}^\nu\sigma_k(\alpha_k q_k + \beta_k p_k )}}{\prod_{k=1}^\nu \cosh \omega_k},
    \
    \sigma_k := \sqrt{2\tanh \omega_k},
\end{equation}
for any $\omega_1, \ldots, \omega_\nu\> 0$, where
\begin{equation}
\label{sk}
  s_k := r_k^\rT r_k = q_k^2 + p_k^2,
  \qquad
  k = 1, \ldots, \nu,
\end{equation}
are pairwise commuting ($[s_j,s_k]=0$ for all $j$, $k$) positive semi-definite self-adjoint quantum variables associated with the vectors $r_k$ in (\ref{r_rk}), and the classical expectation $\bM(\cdot)$ is over mutually independent  standard normal random variables $\alpha_1, \beta_1, \ldots, \alpha_\nu, \beta_\nu$.

\begin{lem}
\label{lem:rand}
Under the conditions of Lemma~\ref{lem:posmom},   the quadratic-exponential function of the quantum variables $X_1, \ldots, X_n$ in (\ref{Xi}) can be represented as
\begin{equation}
\label{Qexp}
  \re^{\frac{\mu}{2} Q}
  =
  \frac{1}{\sqrt{\det \cos(\mu\Theta)}}
  \bM \re^S.
\end{equation}
Here,
\begin{equation}
\label{Sig}
    S
    :=
  \sum_{k=1}^{\nu}
  \sigma_k(\alpha_k q_k + \beta_k p_k)
\end{equation}
is a self-adjoint quantum variable which depends parametrically on the risk-sensitivity parameter $\mu>0$ and the eigenfrequencies $\theta_k$ of the CCR matrix $\Theta$ in (\ref{XCCR}) through
\begin{equation}
\label{uvvarsk}
    \sigma_k
    :=
    \sqrt{2\tanh (\mu\theta_k)},
    \qquad
    k = 1, \ldots, \nu,
\end{equation}
and the classical expectation  $\bM(\cdot)$ is over the
mutually independent  standard normal random variables $\alpha_k$, $\beta_k$.
\end{lem}
\begin{pf}
Due to orthogonality of the matrix $\sqrt{2}H$ in (\ref{eig2_ort2}),  it follows from (\ref{Xqp}), (\ref{QXX}) that
\begin{align}
\nonumber
    Q
  & =
    4
    r^\rT
    (\sqrt{\Gamma}\ox I_2
    )
    H^\rT
    H
    (\sqrt{\Gamma}\ox I_2
    )
    r\\
  \label{Qqp}
    & =
    2
    r^\rT
    (\Gamma\ox I_2
    )
    r
     =
    2
    \sum_{k=1}^{\nu}
    \theta_k s_k,
\end{align}
where use is made of the vector $r$ given by  (\ref{r_rk}), the matrix $\Gamma$ from (\ref{diag}),   and the quantum variables $s_1, \ldots, s_\nu$ from (\ref{sk}).
A combination of (\ref{Qqp}) with the corollary (\ref{qefrand1}) of Lemma~\ref{lem:fact}
yields
\begin{align}
\nonumber
    \re^{\frac{\mu}{2}Q}
    & =
    \prod_{k=1}^{\nu}
    \re^{\mu \theta_k s_k}
    =
    \prod_{k=1}^{\nu}
    \Big(
    \frac{1}{\cosh (\mu \theta_k)}
    \bM
    \re^{\sigma_k(\alpha_k q_k + \beta_k p_k)}
    \Big)\\
\label{eprod}
    & =
    \re^{-\sum_{k=1}^{\nu} \ln \cosh (\mu\theta_k)}
    \bM
    \re^{\sum_{k=1}^{\nu} \sigma_k(\alpha_k q_k + \beta_k p_k) },
\end{align}
which leads to (\ref{Qexp}) in view of (\ref{Sig}), (\ref{uvvarsk}). Here, use is  made of
the relations
$
    \sum_{k=1}^{\nu}\ln \cosh (\mu\theta_k)
  =
  \sum_{k=1}^{\nu}\ln \cos(i \mu \theta_k)
  =
  \frac{1}{2}
  \Tr \ln\cos (\mu \Theta)
  =
  \frac{1}{2}
  \ln\det\cos (\mu \Theta)
$,
where the symmetry of the  functions $\cos$, $\cosh$ and the double multiplicity of the eigenfrequencies $\theta_k$ of the CCR matrix $\Theta$ (as in (\ref{symphiT_phiTtrace})) are taken into account  along with the identity $\Tr \ln N = \ln \det N$ for nonsingular square matrices $N$. The last equality in (\ref{eprod}) uses the commutativity between the position-momentum pairs $(q_k, p_k)$ for different $k$, in accordance with (\ref{rrcomm}),  and the mutual independence of the random variables $\alpha_k$, $\beta_k$.
\hfill$\blacksquare$
\end{pf}

The randomised representation (\ref{Qexp})  of the quadratic-exponential function of the  quantum variables $X_1, \ldots, X_n$ involves only their commutation structure and holds regardless of a particular quantum state.

The classical averaging in Lemma~\ref{lem:rand} over the standard normal random coefficients $\alpha_k$, $\beta_k$ for the positions and momenta $q_k$, $p_k$ in (\ref{Sig}) can be represented in terms of the original quantum variables $X_1, \ldots, X_n$ and the averaging over a classical Gaussian random vector. In what follows, the quantum variables are assumed to have an everywhere finite MGF
\begin{equation}
\label{MGF}
  \Psi(u):= \bE \re^{u^\rT X},
  \qquad
  u \in \mR^n.
\end{equation}
Note that $\Psi$ takes values
in $(0,+\infty)$ (with $\Psi(0) = \bE \cI_\fH=1$) since $\re^{u^\rT X}$  is a positive definite self-adjoint  quantum variable,  which corresponds  formally to the Weyl operator (\ref{cW}) evaluated at a purely imaginary vector $-iu\in i\mR^n$ (that is,  $\re^{u^\rT X} = \cW(-iu)$). Accordingly, (\ref{MGF}) is formally related by $\Psi(u) = \Phi(-iu)$ to  the quasi-characteristic function (QCF) 
\begin{equation}
\label{QCF}
  \Phi(u):= \bE \cW(u) = \overline{\Phi(-u)},
  \qquad
  u \in \mR^n.
\end{equation}
The latter takes values in the disk $\{z\in \mC: |z|\< 1\}$ (with $\Phi(0)=1$) and, due to the Weyl CCRs (\ref{Weyl}),    satisfies a weighted positive semi-definiteness property in the sense that
$    (
        \re^{iu_j^{\rT}\Theta u_k}\Phi(u_j-u_k)
    )_{1\< j,k\< m}
    \succcurlyeq 0
$
for any positive integer $m$ and any points $u_1, \ldots, u_m \in \mR^n$, as a quantum mechanical counterpart  (\cite{CH_1971,H_2010}) of
the Bochner-Khinchin positiveness criterion (\cite{GS_2004})  for the characteristic functions of classical probability distributions.
In contrast to the QCF $\Phi$ in (\ref{QCF}), which is well-defined in $\mR^n$ due to boundedness of the Weyl operators in (\ref{cW}), the finiteness of the MGF $\Psi$ in (\ref{MGF}) everywhere in $\mR^n$ has to be assumed.
The Weyl CCRs (\ref{Weyl}) imply (see also 
\cite[Section~6]{V_2015} or \cite[Lemma~3.1]{VPJ_2019}) that
\begin{equation}
\label{fact}
    \rprod_{k=1}^n
    \re^{u_k X_k}
    =
    \re^{u^\rT X + \frac{i}{2} u^\rT \Theta^\diam u},
    \quad
    u:= (u_k)_{1\< k\< n} \in \mR^n,
\end{equation}
where $\rprod_{k=1}^n \zeta_k := \zeta_1 \x \ldots \x \zeta_n$ is the rightward product of noncommuting variables $\zeta_1, \ldots, \zeta_n$, and
$\Theta^\diam $ denotes a symmetric matrix of order $n$ which inherits its upper triangular part (with the zero main diagonal) from $\Theta$. Accordingly, (\ref{fact}) allows the mixed moments of the quantum variables $X_1, \ldots, X_n$ to be computed in terms of (\ref{MGF}) as
\begin{equation}
\label{moms}
    \bE
    \rprod_{k=1}^n
    X_k^{m_k}
    =
    \d_u^m
    (\Psi(u)
    \re^{\frac{i}{2} u^\rT \Theta^\diam u}
    )\big|_{u=0}
\end{equation}
for any $m:= (m_k)_{1\< k \< n} \in \mZ_+^n$ (with $\mZ_+:= \{0,1,2,\ldots\} $ the set of nonnegative integers, and $\d_u^m:=   \d_{u_1}^{m_1}\ldots \d_{u_n}^{m_n}$ the multiindex notation),
provided $\Psi$ is continuously differentiable at the origin a sufficient number of times.  In the limiting commutative  case, when the CCR matrix $\Theta$ vanishes, so also does  the correction term $\frac{i}{2} u^\rT \Theta^\diam u$ in (\ref{fact}), (\ref{moms}), thus leading to the usual relations  for the mixed moments of classical random variables and their MGF.

\begin{thm}
\label{th:Z}
Under the conditions of Lemmas~\ref{lem:posmom} and \ref{lem:rand}, the QEM  (\ref{Xi}) of the  quantum variables $X_1, \ldots, X_n$ is related to their MGF (\ref{MGF}) as
\begin{equation}
\label{XiPsi}
  \Xi(\mu)
     =
  \frac{  \bM
  \Psi(\sqrt{\mu} Z)}{\sqrt{\det \cos(\mu\Theta)}}.
\end{equation}
Here, the classical expectation $\bM(\cdot)$ is over an auxiliary  zero-mean Gaussian random vector $Z$ in $\mR^n$ with the covariance matrix
\begin{equation}
\label{K}
    K(\mu)
    :=
    \tanc (\mu\Theta),
\end{equation}
which involves the CCR matrix $\Theta$ in (\ref{XCCR}) and the risk sensitivity parameter $\mu>0$
and is a positive definite contraction:
\begin{equation}
\label{K01}
  0 \prec K(\mu) \prec I_n
\end{equation}
(the functions $\tanc z := \frac{\tan z}{z}$, extended  by continuity to $1$ at $z=0$,  and $\cos$ are evaluated (\cite{H_2008})  here at the matrix $\mu\Theta$).
\end{thm}
\begin{pf}
Substitution of (\ref{qk}), (\ref{pk}) into (\ref{Sig}) leads to
\begin{equation}
\label{SZ}
    S
    =
  \sum_{k=1}^{\nu}
    \frac{\sigma_k}{\sqrt{\theta_k}}
  (\alpha_k
  u_k
  +
  \beta_k
  v_k
  )^\rT X
  =
    \sqrt{\mu}
    Z^\rT X,
\end{equation}
where, in view of (\ref{huv}), (\ref{uvvarsk}),
\begin{align}
\nonumber
    Z
    & :=
    \sum_{k=1}^{\nu}
    \sqrt{2\tanhc (\mu\theta_k)}
    (\alpha_k u_k
    +
    \beta_k v_k
    )\\
\nonumber
     & =
    \sum_{k=1}^{\nu}
    \sqrt{2\tanhc (\mu\theta_k)}\,
    h_k
    \gamma_k\\
\label{Z}
    & =
    \sqrt{2}H
    (
    \sqrt{\tanhc (\mu\Gamma)}\ox I_2
    )
    \gamma
\end{align}
(with $\tanhc z := \tanc (-iz)$ a hyperbolic version  of $\tanc$)
is an auxiliary  random vector in $\mR^n$.
Here, $\gamma$ is an $\mR^n$-valued Gaussian random vector, formed from the mutually independent standard normal random variables $\alpha_1, \beta_1, \ldots, \alpha_\nu, \beta_\nu$ as
$  \gamma:= (\gamma_k)_{1\< k \< \nu}$,
with $
  \gamma_k
  :=
  {\scriptsize\begin{bmatrix}
    \alpha_k\\
    \beta_k
  \end{bmatrix}}$ for all
$
  k = 1, \ldots, \nu
$,
and hence, satisfying $\bM \gamma = 0$, $\bM (\gamma\gamma^\rT ) = I_n$. Therefore, $Z$ in (\ref{Z}) is   a zero-mean Gaussian random vector in $\mR^n$ with the covariance matrix
\begin{align}
\nonumber
    &K(\mu) :=
    \bM(ZZ^\rT)\\
\nonumber
    & =
    2H
    (
    \sqrt{\tanhc (\mu\Gamma)}\ox I_2
    )
    \bM (\gamma\gamma^\rT)
    (
    \sqrt{\tanhc (\mu\Gamma)}\ox I_2
    )
    H^\rT\\
\label{MZZ}
    & =
    2H
    (
    \tanhc (\mu\Gamma)\ox I_2
    )
    H^\rT
    =
    \tanc (\mu\Theta),
\end{align}
where the last equality is obtained by applying  (\ref{symphiT_phiTtrace}) to the symmetric function $f(z):= \tanc(\mu z)$.  Since $\tanhc$ on the punctured real line $\mR\setminus \{0\}$ takes values in the interval $(0, 1)$, then so also do the eigenvalues $\tanhc(\mu\theta_k)$   (of double multiplicity) of the matrix $K(\mu)$ in (\ref{MZZ}), which implies (\ref{K01}). Now, (\ref{SZ}) allows (\ref{Qexp}) to be represented as
\begin{align}
\nonumber
  \re^{\frac{\mu}{2} Q}
  & =
  \frac{1}{\sqrt{\det \cos(\mu\Theta)}}
  \bM \re^{\sqrt{\mu} Z^\rT X}\\
\label{Qexp1}
  & =
  \frac{(2\pi)^{-\nu}}{\sqrt{\det (K(\mu)\cos(\mu\Theta))}}
  \int_{\mR^n}
  \re^{\sqrt{\mu} z^\rT X - \frac{1}{2}\|z\|_{K(\mu)^{-1}}^2}
  \rd z,
\end{align}
where $\|v\|_N:= \sqrt{v^\rT N v}$ is a weighted Euclidean (semi-)norm of a real vector $v$  associated with  a real positive (semi-)definite symmetric matrix $N$.
Therefore, by applying the quantum expectation $\bE(\cdot)$ from  (\ref{bE}) to both parts of (\ref{Qexp1}), it follows that
\begin{align}
\nonumber
  \Xi(\mu)
  & =
  \frac{\bE \bM \re^{\sqrt{\mu} Z^\rT X}}{\sqrt{\det \cos(\mu\Theta)}}\\
\nonumber
  & =
  \frac{(2\pi)^{-\nu}}{\sqrt{\det (K(\mu)\cos(\mu\Theta))}}
  \int_{\mR^n}
  \re^{- \frac{1}{2}\|z\|_{K(\mu)^{-1}}^2}
  \bE \re^{\sqrt{\mu} z^\rT X}
  \rd z\\
\label{Qexp2}
  & =
  \frac{  \bM
  \Psi(\sqrt{\mu} Z)}{\sqrt{\det \cos(\mu\Theta)}},
\end{align}
thus establishing (\ref{XiPsi}).   Here,
the commutativity between the classical averaging $\bM(\cdot)$ over $Z$ and the quantum expectation $\bE(\cdot)$ over $X$ (in the sense that $\bE \bM f(X,Z) = \bM \bE f(X,Z)$  for an operator-valued function $f(X,Z)$ of $X$ which depends parametrically on $Z$)  is combined with (\ref{Xi}), (\ref{MGF}).
\hfill$\blacksquare$
\end{pf}

The general structure of the right-hand side of (\ref{Qexp2}) resembles the change of classical Gaussian distributions under translations in the Cameron-Martin theorem (\cite{LS_1977}).  However, the dependence of the  factor $1/\sqrt{\det \cos(\mu\Theta)}$ and the
 covariance matrix (\ref{K})  of the random vector $Z$ on the CCR matrix $\Theta$ is a consequence of noncommutativity of the quantum variables $X_1, \ldots, X_n$. More precisely, this dependence originates from the Weyl CCRs (\ref{Weyl}) (see the proof of Lemma~\ref{lem:fact}  in \cite[Theorem A.1]{VPJ_2021}). In the limiting
classical case, when $\Theta$ vanishes (together with its eigenfrequencies $\theta_1, \ldots, \theta_\nu$), the covariance matrix $K(\mu)$ in (\ref{K}) reduces to $I_n$,  and $Z$ becomes a standard normal random vector in $\mR^n$ in accordance with (\ref{MMM}).

\section{QEMs for Gaussian states and a non-Gaussian upper bound}
\label{sec:Gauss}

The quantum state $\rho$ enters the randomised representation of the QEM  in (\ref{XiPsi})  through the MGF (\ref{MGF}) of the quantum variables $X_1, \ldots, X_n$ whose commutation structure specifies the covariance matrix (\ref{K}) of the auxiliary zero-mean Gaussian random vector $Z$.
If the quantum variables $X_1, \ldots, X_n$ are in a Gaussian state (\cite{KRP_2010}) with the mean vector $M:= \bE X \in \mR^n$ and quantum covariance matrix
\begin{equation}
\label{covX}
    \cov(X):= \bE((X-M)(X-M)^\rT) = C + i\Theta \succcurlyeq 0,
\end{equation}
where the imaginary part $\Im \cov(X) = \Theta$ does not depend on a particular state, and  $C:= \Re \cov(X) = C^\rT \succcurlyeq 0$ (as the real part of a positive semi-definite Hermitian matrix), then the MGF (\ref{MGF}) takes the form
\begin{equation}
\label{PsiGauss}
    \Psi(u) = \re^{M^\rT u + \frac{1}{2}\|u\|_C^2},
    \qquad
    u \in \mR^n.
\end{equation}
This is identical to the MGF of the corresponding  classical Gaussian distribution in $\mR^n$ except for the positive semi-definiteness in (\ref{covX}), which is a stronger condition than $C \succcurlyeq 0$ and originates from the Heisenberg uncertainty relation (\cite{H_2001}). In the Gaussian quantum state,  the randomised representation (\ref{XiPsi}) allows the QEM to be computed in closed form. To this end, we will use the identity
\begin{equation}
\label{gint}
    (2\pi)^{-m/2}
  \sqrt{\det N}
  \int_{\mR^m}
  \re^{a^\rT u - \frac{1}{2}\|u\|_N^2}
  \rd u
  =
  \re^{\frac{1}{2}\|a\|_{N^{-1}}^2},
\end{equation}
which holds for any $a \in \mR^m$ and positive definite matrix $N=N^\rT \in \mR^{m\x m}$ and describes the MGF of a zero-mean Gaussian distribution in $\mR^m$ with the precision matrix $N$.

\begin{thm}
\label{th:QEMGauss}
In addition to the conditions of Theorem~\ref{th:Z}, suppose the  quantum variables $X_1, \ldots, X_n$ are in a Gaussian state with the mean vector $M$ and the quantum covariance matrix (\ref{covX}). Then the  logarithm (\ref{Ups}) of their QEM  (\ref{Xi}) is computed as
\begin{align}
\nonumber
  \Ups(\mu)
  = &
  \frac{1}{2}
  \big(
    \|M\|_{\mu K(\mu)(I_n-\mu CK(\mu))^{-1}}^2\\
\label{UpsGauss}
    & -
    \ln \det (\cos (\mu \Theta) - \mu C \sinc(\mu \Theta))
  \big)
\end{align}
(with $\sinc z := \frac{\sin z}{z}$ extended  by continuity to $1$ at $z=0$), provided the risk  sensitivity  parameter $\mu>0$ satisfies
\begin{equation}
\label{mu*}
  \mu \br(CK(\mu))<1,
\end{equation}
where $\br(\cdot)$ is the spectral radius. \hfill$\square$
\end{thm}
\begin{pf}
In application to (\ref{PsiGauss}), the classical expectation in (\ref{XiPsi}) takes the form
\begin{align}
\nonumber
  \bM &\Psi(\sqrt{\mu}Z)
   \!=\!
  \frac{(2\pi)^{-\nu}}{\sqrt{\det(\mu K(\mu))}}
  \int_{\mR^n}\!\!
  \re^{M^\rT u - \frac{1}{2} \|u\|_{(\mu K(\mu))^{-1}-C}^2}
  \rd u\\
\label{bMPsi}
  & =
  \frac{1}{\sqrt{\det(I_n - \mu C K(\mu))}}
  \re^{\frac{1}{2} \|M\|_{\mu K(\mu)(I_n-\mu CK(\mu))^{-1}}^2},\!\!\!
\end{align}
where (\ref{gint}) is used with $a:= M$ and $N:= (\mu K(\mu))^{-1}-C =(I_n - \mu  C K(\mu))(\mu K(\mu))^{-1} \succ 0$ under the condition (\ref{mu*}).  By substituting (\ref{bMPsi}) into (\ref{XiPsi}) and using the identity  $\tanc z \cos z = \sinc z$, whereby
\begin{equation}
\label{Kcos}
    K(\mu)\cos(\mu \Theta) = \sinc(\mu\Theta)
\end{equation}
in view of (\ref{K}), it follows that
\begin{align}
\nonumber
  \Xi(\mu)
     & =
  \frac{\re^{\frac{1}{2} \|M\|_{\mu K(\mu)(I_n-\mu CK(\mu))^{-1}}^2}}{\sqrt{\det((I_n - \mu C K(\mu)) \cos(\mu\Theta))}}
  \\
\label{XiPsi1}
  & =
  \frac{\re^{\frac{1}{2} \|M\|_{\mu K(\mu)(I_n-\mu CK(\mu))^{-1}}^2}}{\sqrt{\det(\cos(\mu\Theta) - \mu C \sinc(\mu\Theta) )}}
  .
\end{align}
Taking the logarithm on both sides of (\ref{XiPsi1}) establishes (\ref{UpsGauss}) for (\ref{Ups}).
\hfill$\blacksquare$
\end{pf}

In the limiting
classical case of $\Theta = 0$, mentioned at the end of Section~\ref{sec:rand}, with $K(\mu)=I_n$ in (\ref{K}),   the condition (\ref{mu*}) reduces to $\mu < 1/\br(C)$, and  the representation (\ref{UpsGauss}) takes the form
$
    \Ups(\mu) = \frac{1}{2}\big(\|M\|_{\mu (I_n-\mu C)^{-1}}^2-\ln \det (I_n - \mu C )\big)
$, which plays a part in classical risk-sensitive  control (\cite{J_1973,W_1981}).
Returning to the quantum case, since the Gaussian MGF (\ref{PsiGauss}) has a quadratic-exponential growth at infinity,
its weighted $L_2$-norm (\cite{V_2015})
\begin{equation}
\label{Psinorm}
    \sn \Psi\sn_P
    :=
    \sqrt{\int_{\mR^n}\re^{-\|u\|_P^2} \Psi(u)^2\rd u},
\end{equation}
specified by a real positive definite symmetric matrix $P$ of order $n$,  is finite whenever $P \succ C$.  Moreover, it admits a closed-form computation (by using (\ref{gint}) with $a:= 2M$ and $N:= 2(P-C)$  similarly to (\ref{bMPsi})) as 
     $\sn \Psi\sn_P
      =
    \sqrt{\int_{\mR^n}\re^{2M^\rT u-\|u\|_{P-C}^2}\rd u}
    =
    \frac{\pi^{\nu/2} \re^{\frac{1}{2}\|M\|_{(P-C)^{-1}}^2}}{\sqrt[4]{\det (P-C)}}$. 
For a general (not necessarily Gaussian) quantum state,
the norms (\ref{Psinorm}), considered for an interval of weighting matrices $P$,    lead   to the following upper bound on the QEM in (\ref{Xi}).

\begin{thm}
\label{th:upper}
Under the conditions of Theorem~\ref{th:Z},  the QEM (\ref{Xi}) of the quantum variables $X_1, \ldots, X_n$ admits un upper bound
\begin{equation}
\label{Xiupper}
    \Xi(\mu)
    \!\!\< \!
\frac{(4\pi)^{-\nu/2}}{\sqrt{\det(\mu \sinc(\mu\Theta))}}\!
\inf_{0\prec P \prec \frac{1}{\mu}K(\mu)^{-1}}\!\!
    \frac{\sn \Psi\sn_P}{\sqrt[4]{\det (\frac{1}{\mu} K(\mu)^{-1}\!-\!\!P)}}
\end{equation}
in terms of the  weighted norms (\ref{Psinorm}) of the MGF (\ref{MGF}),
where the matrix $K(\mu)$ is given by (\ref{K}).
\end{thm}
\begin{pf}
A combination of (\ref{XiPsi}) with the Cauchy-Bunyakov\-sky-Schwarz inequality (for the weighted inner product $\dbra f, g\dket_P :=
\int_{\mR^n} \re^{-\|u\|_P^2} \overline{f(u)} g(u)\rd u$ of functions $f,g: \mR^n \to \mC$ which generates the norm (\ref{Psinorm})) yields
\begin{align}
\nonumber
  &\Xi(\mu)
     =
  \frac{(2\pi)^{-\nu}}{\sqrt{\det(\mu K(\mu) \cos(\mu\Theta))}}
  \int_{\mR^n}
  \re^{-\frac{1}{2}\|u\|_{(\mu K(\mu))^{-1}}^2}
  \Psi(u)
  \rd u\\
\nonumber
     & =
  \frac{(2\pi)^{-\nu}}{\sqrt{\det(\mu \sinc(\mu\Theta))}}
  \!\int_{\mR^n}\!
  \re^{-\frac{1}{2}\|u\|_{(\mu K(\mu))^{-1}-P}^2}
  \re^{-\frac{1}{2}\|u\|_P^2}
  \Psi(u)
  \rd u\\
\nonumber
    & \<
  \frac{(2\pi)^{-\nu}}{\sqrt{\det(\mu \sinc(\mu\Theta))}}
  \sqrt{
  \int_{\mR^n}
  \re^{-\|u\|_{(\mu K(\mu))^{-1}-P}^2}
  \rd u }\,
  \sn \Psi\sn_P\\
\label{Xiupper1}
    & =
  \frac{(4\pi)^{-\nu/2}}{\sqrt{\det(\mu \sinc(\mu\Theta))}}
    \frac{\sn \Psi\sn_P}{\sqrt[4]{\det ((\mu K(\mu))^{-1}-P)}},
\end{align}
where use is made of (\ref{Kcos}), and
$P=P^\rT \in \mR^{n\x n}$ is an arbitrary matrix satisfying
\begin{equation}
\label{Pint}
    0\prec P \prec \frac{1}{\mu}K(\mu)^{-1}.
\end{equation}
The second inequality in (\ref{Pint})  secures finiteness of the third integral in (\ref{Xiupper1}) (which can be  computed by using (\ref{gint}) with $a:= 0$ and $N:=2((\mu K(\mu))^{-1}-P)\succ 0$).
Since the left-hand side of (\ref{Xiupper1}) does not depend on $P$, this upper bound can be tightened by taking the infimum on its right-hand side over the matrix interval (\ref{Pint}), which leads to (\ref{Xiupper}) in view of (\ref{Kcos}).
\hfill$\blacksquare$
\end{pf}

A simplified upper bound is obtained from (\ref{Xiupper}) by restricting the set of weighting matrices $P$ to scalar matrices $P=\lambda I_n$, where it is assumed that
\begin{align}
\nonumber
    0 < \lambda &< \frac{1}{\mu} \lambda_{\min}(K(\mu)^{-1}) = \frac{1}{\mu \lambda_{\max}(K(\mu))}\\
\label{lam*}
     & = \frac{1}{\mu \tanhc (\mu\min_{1\< k \< \nu} \theta_k)}
    =:
    \lambda_*(\mu)
\end{align}
in order to satisfy (\ref{Pint}),
with $\lambda_{\min}(\cdot)$, $\lambda_{\max}(\cdot)$ the smallest and largest eigenvalues. 
In (\ref{lam*}), the spectrum $\{\tanhc(\mu\theta_k): k= 1, \ldots, \nu\}$ (as the set of eigenvalues, ignoring their double multiplicity) of the matrix $K(\mu)$ from (\ref{K}) is used along with the property that the symmetric function $\tanhc$ is strictly decreasing on $\mR_+$. Accordingly, the inequality (\ref{Xiupper}) is replaced in this case with its weaker counterpart
\begin{equation}
\label{Xiupper2}
    \Xi(\mu)
    \!\!\< \!
\frac{(4\pi)^{-\nu/2}}{\sqrt{\det(\mu \sinc(\mu\Theta))}}\!
    \inf_{0< \lambda <\lambda_*(\mu)}\!
    \frac{\sn \Psi\sn_\lambda}{\sqrt[4]{\det (\frac{1}{\mu}K(\mu)^{-1}\!-\!\!\lambda I_n)}},
\end{equation}
where the second denominator involves the characteristic polynomial of the matrix $\frac{1}{\mu} K(\mu)^{-1}$, and, with a slight abuse of notation,  a shorthand
\begin{equation}
\label{Psinormlam}
    \sn \Psi\sn_\lambda
    :=
    \sqrt{\int_{\mR^n}\re^{-\lambda |u|^2} \Psi(u)^2\rd u},
    \qquad
    \lambda >0,
\end{equation}
is used for (\ref{Psinorm})  in the case of scalar weighting matrices $P$. Note that $\sn \Psi \sn_{\lambda_{\max}(P)}\< \sn \Psi \sn_P \<  \sn \Psi \sn_{\lambda_{\min}(P)}$ for any $P\succ 0$. The numerator and denominator of the fraction, which  is minimised in (\ref{Xiupper2}), are strictly decreasing functions of $\lambda$ over the interval (\ref{lam*}), thus making the infimum nontrivial.

\section{QEM bounds for open quantum harmonic oscillators}
\label{sec:OQHO}

As an application of the above results to probabilistic bounds for quantum stochastic systems, suppose the underlying system is a $\nu$-mode open quantum harmonic oscillator (OQHO) (\cite{GZ_2004}),  with its dynamic  variables evolving in time $t\> 0$ according to a linear Hudson-Parthasarathy quantum stochastic differential equation (QSDE) (\cite{P_1992}):
\begin{equation}
\label{dXlin}
    \rd X = A X\rd t + B\rd W
\end{equation}
(the time arguments are omitted).
It is driven by an $\frac{m}{2}$-channel  quantum Wiener process $W:= (W_k)_{1\< k \< m}$ as an external bosonic field (with an even number $m$ of time-varying self-adjoint components $W_1, \ldots, W_m$) on a symmetric Fock space $\fF$ with a quantum Ito table $\rd W \rd W^\rT = (I_m + iJ)\rd t$ (so that  $[\rd W, \rd W^\rT] = 2i J \rd t$), where  $J:= I_{m/2}\ox \bJ$, with $\bJ$ from (\ref{bJ}). The matrices
\begin{equation}
\label{AB}
    A:= 2\Theta (R + N^{\rT}JN), 
    \qquad
    B:= 2\Theta N^{\rT}
\end{equation}
in (\ref{dXlin}) are parameterised by an  energy matrix $R = R^\rT \in \mR^{n\x n}$ and a coupling matrix $N \in \mR^{m\x n}$ which specify the Hamiltonian $\frac{1}{2}X^\rT R X$ and the vector $NX$ of $m$ operators of coupling of the system to the external fields. Accordingly, $\fH = \fH_0 \ox \fF$ is the system-field  tensor-product space, with $\fH_0$  the initial system space. The density operator in (\ref{bE}) is assumed to be $\rho := \rho_0 \ox \ups$, where $\rho_0$ is the initial system state on $\fH_0$, and $\ups$ is the vacuum field state on the Fock space $\fF$. Then, due to linearity of the QSDE (\ref{dXlin}), the MGF $\Psi_t$  for the system variables at time $t\> 0$ in (\ref{MGF}) is related to the initial MGF $\Psi_0$ by
\begin{align}
\nonumber
    \Psi_t(u)
    & := \bE \re^{u^\rT X(t)}
    =
    \bE \re^{u^\rT \re^{tA}X(0)}
    \bE \re^{u^\rT \int_0^t \re^{(t-s)A} B\rd W(s)}    \\
\label{Psit}
    & = \Psi_0(\re^{tA^\rT}u)\re^{\frac{1}{2}\|u\|_{\Sigma_t}^2},
    \qquad
    u \in \mR^n
\end{align}
(cf. \cite[Eq. (108)]{V_2015}),   where the matrix
\begin{equation}
\label{Sigmat}
    \Sigma_t
    :=
    \int_0^t
    \re^{sA} BB^\rT \re^{sA^\rT} \rd s,
    \qquad
    t\> 0,
\end{equation}
is the finite-horizon controllability Gramian for the pair $(A,B)$.  If the initial system state is Gaussian, then so also are its subsequent states (\cite{VPJ_2018}) and their QEMs can be computed in closed form according to Theorem~\ref{th:QEMGauss}. However, (\ref{Psit}) is valid for non-Gaussian states as well and allows norm bounds for the MGF $\Psi_t$ to be obtained in terms of those for $\Psi_0$, with the subsequent application of these bounds to the QEMs.

\begin{thm}
\label{th:Psi}
Suppose the initial system variables of the OQHO (\ref{dXlin}) with vacuum input fields have an everywhere finite MGF $\Psi_0$. Then the norms  (\ref{Psinormlam}) of the subsequent MGFs $\Psi_t$ in (\ref{Psit}) are related by
\begin{equation}
\label{PsiPsi}
    \sn \Psi_t\sn_{\lambda}
    =
    \re^{-\frac{t}{2}\Tr A}
    \sn \Psi_0\sn_{\Pi_{t,\lambda}},
    \qquad
    t \> 0,
\end{equation}
to the norm (\ref{Psinorm}) of the initial MGF with the weighting matrix
\begin{equation}
\label{Pit}
    \Pi_{t,\lambda}:= \re^{-tA}(\lambda I_n - \Sigma_t)\re^{-tA^\rT}\succ 0
\end{equation}
associated with the controllability Gramian (\ref{Sigmat}), provided
\begin{equation}
\label{lam}
    \lambda > \lambda_{\max}(\Sigma_t).
\end{equation}
\end{thm}
\begin{pf}
Application of (\ref{Psinormlam}) to the MGF $\Psi_t$ in (\ref{Psit}) yields
\begin{align}
\nonumber
    \sn \Psi_t\sn_\lambda^2
     & =
    \int_{\mR^n}
    \re^{\|u\|_{\Sigma_t}^2-\lambda |u|^2}
    \Psi_0(\re^{tA^\rT}u)^2
    \rd u\\
\nonumber
    & =
    \re^{-t\Tr A}
    \int_{\mR^n}
    \re^{- \|\re^{-tA^\rT}v\|_{\lambda I_n - \Sigma_t}^2}
    \Psi_0(v)^2
    \rd v\\
\label{Psitnorm}
    & =
    \re^{-t\Tr A}
    \sn \Psi_0\sn_{\Pi_{t,\lambda}}^2,
\end{align}
which establishes (\ref{PsiPsi}).
The new integration variable $v:= \re^{tA^\rT} u$ in (\ref{Psitnorm})   (with the Jacobian $\det \d_v u = \det \re^{-tA^\rT} = \re^{-t\Tr A}$) is used along with the norm (\ref{Psinorm}) specified by the matrix (\ref{Pit}) whose positive definiteness is secured by the condition (\ref{lam}).
\hfill$\blacksquare$
\end{pf}

The right-hand side of (\ref{lam}) is a nondecreasing function of time,  which has a finite limit $\lim_{t\to +\infty}\lambda_{\max}(\Sigma_t) = \lambda_{\max}(\Sigma_\infty)$ if the matrix $A$ in (\ref{AB}) is Hurwitz, where
$    \Sigma_\infty
    :=
    \int_0^{+\infty}
    \re^{sA} BB^\rT \re^{sA^\rT} \rd s
$
is the infinite-horizon controllability Gramian of the pair $(A,B)$. If the risk sensitivity parameter $\mu>0$ is small enough, so that the quantity $\lambda_*(\mu)$ in (\ref{lam*}) satisfies
$
    \lambda_*(\mu) > \lambda_{\max}(\Sigma_\infty)
$, then the interval $\Delta_t := (\lambda_{\max}(\Sigma_t), \lambda_*(\mu))$ is nonempty for any time $t\> 0$.  In this case, a combination of (\ref{Xiupper2}) with (\ref{PsiPsi}) leads to the following upper bound for the QEM of the system variables of the OQHO at time $t\> 0$:
\begin{equation}
\label{Xitupper2}
    \Xi_t(\mu)
    \<
\frac{(4\pi)^{-\nu/2} \re^{-\frac{t}{2}\Tr A}}{\sqrt{\det(\mu \sinc(\mu\Theta))}}
    \inf_{\lambda \in \Delta_t}
    \frac{
    \sn \Psi_0\sn_{\Pi_{t,\lambda}}}{\sqrt[4]{\det (\frac{1}{\mu}K(\mu)^{-1}-\lambda I_n)}}.
\end{equation}
The initial MGF $\Psi_0$ enters this inequality through its weighted $L^2$-norms, thus describing the influence of the (not necessarily Gaussian) initial quantum state of the system variables on their subsequent QEMs,  which can be used for statistical ``localisation'' in terms of the tail probability bounds (\ref{tail}).

Although (\ref{Xitupper2}) takes advantage of the relatively simple MGF evolution (\ref{Psit}) for OQHOs with vacuum input fields, this approach can, in principle, be extended to nonlinear quantum stochastic systems with the Hamiltonian and coupling operators in a Weyl quantization form. In the latter  case, the MGF evolves according to an in\-teg\-ro-differential equation (\cite{V_2015}) which is amenable to the development of relevant dissipation inequalities.

\section{Conclusion}
\label{sec:conc}

For quantum system variables of position-momentum type,  we have discussed a randomised representation  of their QEMs in terms of the MGF.  This has led to upper bounds for the QEMs (applicable to the tail distribution bounds for the quadratic function of the system variables) in terms of the weighted $L^2$-norms of the MGF. This approach reduces the problem of ``localising'' the behaviour of an underlying quantum stochastic system in terms of the QEMs  to the analysis of the norms of the  MGF whose evolution is more tractable. Such analysis can be carried out through dissipation inequalities for the MGF or by using its closed-form dynamics available in the case of linear QSDEs, as has been illustrated for OQHOs with vacuum input fields and a class of (not necessarily Gaussian) initial states.

\appendix
\section{Alternative 
proof of (\ref{MMM})}
\label{sec:MMM}

Since a classical random vector  $X$ in $\mR^n$ satisfies $\sqrt{\mu}z^\rT X - \frac{1}{2}|z|^2 = \frac{\mu}{2} |X|^2 - \frac{1}{2}|z-\sqrt{\mu}X|^2$ for any $\mu\> 0$, $z \in \mR^n$, then, in view of (\ref{nnu}), (\ref{MGFclass}), the right-hand side of (\ref{MMM}) can be  represented as
\begin{align*}
    &(2\pi)^{-\nu}
    \int_{\mR^n}
    \re^{-\frac{1}{2}|z|^2}
    \Lambda(\sqrt{\mu}z)
    \rd z\\
    & =
    (2\pi)^{-\nu}
    \bM
    \int_{\mR^n}
    \re^{\sqrt{\mu}z^\rT X - \frac{1}{2}|z|^2}
    \rd z\\
    & =
    \bM
    \Big(
        \re^{\frac{\mu}{2}|X|^2}
        \underbrace{
        (2\pi)^{-\nu}
        \int_{\mR^n}
        \re^{- \frac{1}{2}|z-\sqrt{\mu}X|^2}
        \rd z}_{1}
    \Big)
    =
    \bM \re^{\frac{\mu}{2}|X|^2},
\end{align*}
which yields the left-hand side of (\ref{MMM}), thus proving  the relation.
                                                                         

\begin{thebibliography}{xx}  % you can also add the bibliography by hand
\bibitem[Accardi \& Boukas(2018)]{AB_2018}
L.Accardi, and A.Boukas, Normally ordered disentanglement of multi-dimensional
Schr\"{o}dinger algebra exponentials, \emph{Comm. Stoch. Anal.},  vol. 12, 2018, pp. 283--328.

\bibitem[Arnold(1989)]{A_1989}
V.I.Arnold, \emph{Mathematical Methods of Classical Mechanics}, 2nd Ed., Springer, New York, 1989.



\bibitem[Boukas(1996)]{B_1996}
A.Boukas, Stochastic control of operator-valued processes in boson Fock space, \emph{Russian
J. Math. Phys.}, vol. 4, 1996, pp. 139--150.

\bibitem[Bregman(1967)]{B_1967}
L.M.Bregman, The relaxation method of finding the common points of convex sets and its application to the solution of problems in convex programming,  \emph{USSR Comput. Math. Math. Phys.}, vol. 7, no. 3, 1967, pp. 200--217.

\bibitem[Cushen \& Hudson(1971)]{CH_1971}
C.D.Cushen, and R.L.Hudson, A quantum-mechanical central limit theorem,
\emph{J. Appl. Prob.}, vol. 8, no. 3, 1971, pp. 454--469.


\bibitem[Folland(1989)]{F_1989}
G.B.Folland, \emph{Harmonic Analysis in Phase Space}, Princeton University Press, Princeton, 1989.

\bibitem[Gardiner \& Zoller(2004)]{GZ_2004}
C.W.Gardiner, and P.Zoller,
\emph{Quantum Noise},
Springer, Berlin, 2004.

\bibitem[Gikhman \& Skorokhod(2004)]{GS_2004}
I.I.Gikhman, and A.V.Skorokhod,
\emph{The Theory of Stochastic Processes}, vol. I, Springer, Berlin,
2004.

\bibitem[Higham(2008)]{H_2008}
N.J.Higham,
\textit{Functions of Matrices},
SIAM, Philadelphia,  2008.



\bibitem[Holevo(2001)]{H_2001}
A.S.Holevo, \emph{Statistical Structure of Quantum Theory}, Springer, Berlin, 2001.


\bibitem[Hudson \& Parthasarathy(1984)]{HP_1984}
R.L.Hudson,  and K.R.Parthasarathy, Quantum Ito’s formula and stochastic evolutions,
\emph{Commun. Math. Phys.}, vol.  93, 1984,  pp. 301--323.

\bibitem[Hudson(2010)]{H_2010}
R.L.Hudson,
Quantum Bochner theorems and incompatible observables,
\emph{Kybernetika}, vol. 46, no. 6, 2010, pp. 1061--1068.

\bibitem[Hudson(2018)]{H_2018}
R.L.Hudson, A short walk in quantum probability, \emph{Philos. Trans. R. Soc. A}, vol.  376,
2018, pp.  1--13.

\bibitem[Jacobson(1973)]{J_1973}
D.H.Jacobson, Optimal stochastic linear systems with exponential performance
criteria and their relation to deterministic differential games, \emph{IEEE Trans. Autom.
Control}, vol. 18, 1973, pp. 124--31.




\bibitem[Liptser \& Shiryayev(1977)]{LS_1977}
R.S.Liptser, and A.N.Shiryayev,
\emph{Statistics of Random Processes I: General Theory},
Springer-Verlag, Berlin, 1977.

\bibitem[Nielsen \& Chuang(2000)]{NC_2000}
M.A.Nielsen, and I.L.Chuang,
\textit{Quantum Computation and Quantum Information},
Cambridge University Press, Cambridge, 2000.

\bibitem[Parthasarathy(1992)]{P_1992}
K.R.Parthasarathy,
\emph{An Introduction to Quantum Stochastic Calculus},
Birk\-h\"{a}user, Basel, 1992.

\bibitem[Parthasarathy(2010)]{KRP_2010}
K.R.Parthasarathy,
What is a Gaussian state?
\emph{Comm. Stoch. Anal.}, vol. 4, no. 2, 2010, pp. 143--160.


\bibitem[Rockafellar(1970)]{R_1970}
R.T.Rockafellar,
\emph{Convex Analysis}.
Princeton University Press, Princeton, New Jersey, 1970.

\bibitem[Sakurai(1994)]{S_1994}
J.J.Sakurai,
\emph{Modern Quantum Mechanics},
 Addison-Wesley, Reading, Mass., 1994.

 \bibitem[Parthasarathy \& Sengupta(2015)]{PS_2015}
 K.R.Parthasarathy,  and R.Sengupta, From particle counting to Gaussian tomography,
\emph{Inf. Dim. Anal. Quantum Prob. Rel. Topics}, vol.  18,  2015, pp.  1550023.


\bibitem[Shiryaev(1996)]{S_1996}
A.N.Shiryaev, \emph{Probability}, 2nd Ed., Springer, New York, 1996.

\bibitem[Vladimirov(2015)]{V_2015}
I.G.Vladimirov, Evolution of quasi-characteristic functions in quantum stochastic systems with Weyl quantization of energy operators,
(arXiv:1512.08751 [math-ph], 29 December 2015).


\bibitem[Vladimirov \& Petersen(2012)]{VP_2012}
I.G.Vladimirov,  and I.R.Petersen, Risk-sensitive dissipativity of linear quantum stochastic systems under Lur'e type perturbations of Hamiltonians, 2nd Australian Control Conference, 15-16 November 2012, Sydney, Australia,  pp. 247--252.

\bibitem[Vladimirov, Petersen \& James(2018)]{VPJ_2018}
I.G.Vladimirov, I.R.Petersen, and M.R.James, Multi-point Gaussian states, quadratic-exponential cost functionals, and large deviations estimates for linear quantum stochastic systems, \textit{Appl. Math. Optim.}, vol. 83, no. 1, 2021, pp. 83--137 (published online 24 July 2018).


 \bibitem[Vladimirov, Petersen \& James(2019)]{VPJ_2019}
I.G.Vladimirov, I.R.Petersen, and M.R.James,
Parametric randomization, complex symplectic factorizations, and quadratic-exponential functionals for Gaussian quantum states,  \emph{Inf. Dim. Anal. Quant. Prob. Rel. Top.}, vol. 22, no. 3, 2019, 1950020.


\bibitem[Vladimirov, Petersen \& James(2021)]{VPJ_2021}
I.G.Vladimirov, I.R.Petersen, and M.R.James,
Quadratic-exponential functionals of Gaussian
quantum processes,
\emph{Inf. Dimen. Anal. Quant. Prob.},
vol. 24, no. 4, 2021,  pp. 2150024-1--40.


\bibitem[Walls \& Milburn(2008)]{WM_2008}
D.F.Walls, and G.J.Milburn,
\emph{Quantum Optics}, 2nd Ed., Springer, Berlin, 2008.

\bibitem[Whittle(1981)]{W_1981}
P.Whittle, Risk-sensitive linear/quadratic/Gaussian control, \emph{Adv. Appl. Probab.},
vol. 13, 1981, pp. 764--77.


\bibitem[Wiseman \& Milburn(2010)]{WM_2010}
H.M.Wiseman, and G.J.Milburn,
\emph{Quantum measurement and control},
Cambridge University Press,
Cambridge, 2010.



\end{thebibliography}
\end{document}